\documentclass[a4paper,compress,fleqn]{cas-dc}

\usepackage[version=4]{mhchem}
\usepackage[running]{lineno}

\usepackage[numbers,sort&compress]{natbib}

\def\tsc#1{\csdef{#1}{\textsc{\lowercase{#1}}\xspace}}
\tsc{WGM}
\tsc{QE}
\tsc{EP}
\tsc{PMS}
\tsc{BEC}
\tsc{DE}

\begin{document}

\shorttitle{Numerical simulation of the response of single gap timing RPCs with the space charge effects and Garfield++}

\shortauthors{Tanay Dey et~al.}

\title [mode = title]{Numerical simulation of the response of single gap timing RPCs with the  space charge effects and Garfield++}                      
			\author[1,2]{Tanay Dey}[orcid=0000-0001-7996-044X]
          \cormark[1]
		\ead{tanay.jop@gmail.com}
		\author[3]{Purba Bhattacharya}
		\author[1,4]{Supratik Mukhopadhyay}
		\author[1,4]{Nayana Majumdar}
        \author[5]{Abhishek Seal}
		\author[2]{Subhasis Chattopadhyay}
	
				\address[1]{Homi Bhabha National Institute,
			Mumbai,
			India}
		\address[2]{Variable Energy Cyclotron centre, Kolkata, India}
		
		\address[3]{Department of Physics, School of Basic and Applied Sciences, Adamas University, Kolkata, India}
		
		\address[4]{Saha Institute of Nuclear Physics, Kolkata, India
		}
		\address[5]{Regent Education and Research Foundation, Kolkata, India}
\cortext[cor1]{Corresponding author}

\begin{abstract}
In this article, we report the simulated response of timing RPCs of different gas gaps. A 3D Montecarlo code was developed and integrated with Garfield++ to simulate the avalanche processes with space charge effects which allow actual charge and timing spectrums. The results of this study are presented with examples of timing RPCs of gas gaps 0.02 cm and 0.03 cm.
\end{abstract}




\maketitle

	\section{Introduction}\label{sec:1_intro}
		 Montecarlo tools play an important role in simulating physics processes in detectors related to high energy physics. These tools allow us to predict signal pulse amplitude and time of threshold crossing, time resolution, efficiency, etc. A realistic detector physics simulation of a resistive plate chamber can be done by considering the dynamic space charge effect. The C++-based software Garfield++ ~\cite{Garfield}  can be used to simulate the detector physics of resistive plate chambers (RPC) \cite{cardeli-1,cardeli-2}. The 3D particle tracing model in Garfield++ is an exact tool to simulate electron avalanches inside the RPC. However, the absence of dynamic space charge calculation in Garfield++ made it a step behind in generating a real avalanche. 
          \par In this work, the main motivation is to discuss the variation of induced charge and time resolution with the gas gap of the RPC. To consider the dynamic space charge field, we have implemented a 3D line charge model inside Garfield++ as discussed in \cite{Dey_2020,Dey_2022}. 
   To test the working of this model, we have taken two numerical timing RPCs of gas gap 0.02 cm and 0.03 cm, and a comparison of the results has been discussed. Another discussion on the simulation of the response of single gap timing RPC with different space charge models can be found in \cite{LIPPMANN200319}. The experimental results of single gap and multigap timing RPCs for the same gas gaps can be found in ~\cite{BLANCO2002328,BLANCO200370,Fonte:491918}.
   \par In section \ref{sec:field} the applied electric field configuration of RPCs has been discussed. The number of primary electron distributions inside the gas gap of two RPCs has been discussed in section \ref{sec:track}. The sections \ref{sec:indCh} and \ref{sec:risetime} discussed a comparison of induced charge and signal rise time for different geometrically configured RPCs at the same applied field. Finally in section \ref{sec:8_summary}
the conclusion or summary  of this study has been discussed.
	\section{Estimation of electric field inside RPC using neBEM}\label{sec:field}
	  
\begin{figure}[h]
	  	\center\includegraphics[scale=0.3]{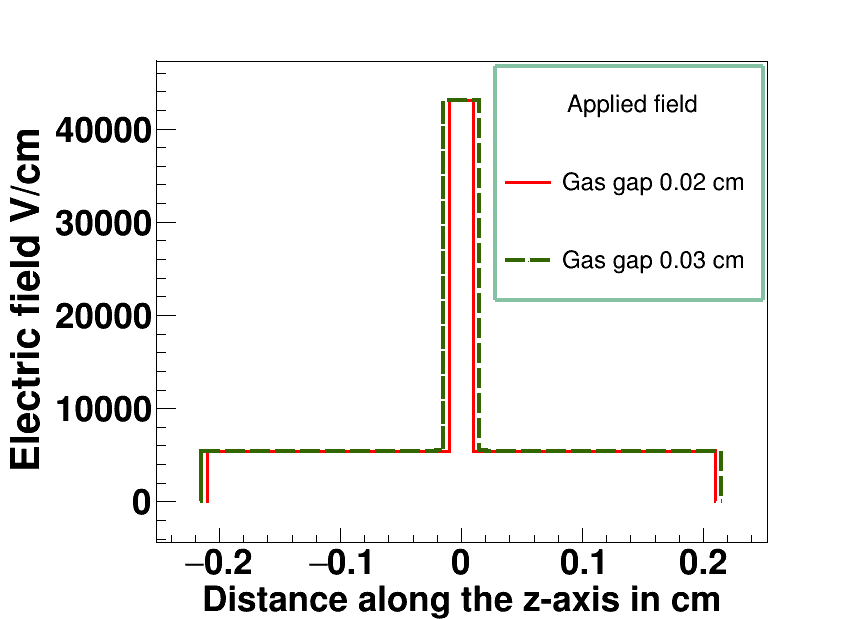}
	  	\caption{Applied field for RPCs of gas-gaps  0.02 cm and 0.03 cm \label{fig:Applied_field}}
	  \end{figure}

Two RPCs of gas-gap 0.02 cm (RPC1) and 0.03 cm (RPC2) have been designed with the help of the geometrical tool of Garfield++. The electrode thickness and area of the surface are considered as 0.2 cm and 30 cm$^2$, respectively. The applied voltages for the RPC1 and RPC2 have been chosen such that the electric field inside each RPCs remains the same (43 kV/cm) as shown in Figure \ref{fig:Applied_field}. The values of the applied voltages on RPC1, and RPC2 are 1505 V and 1720 V respectively. A gas mixture of $\ce{C_2H_2F_4}$ (85\%), $\ce{i\text{-}C_4H_10}$ (5\%), $\ce{SF_6}$ (10\%) has been used in each RPCs. 
 \section{Distribution of primary cluster in RPCs of different gas-gap} \label{sec:track}
	The size of the avalanche in an RPC detector depends on the number of primary electrons. Therefore it is necessary to understand the distribution of the primary electrons in the different gas gaps of the RPCs. The following discussion will discuss the distribution of primary electrons inside RPC1 and RPC2.
 
	\par
\begin{figure}[h]
		\center\includegraphics[scale=0.3]{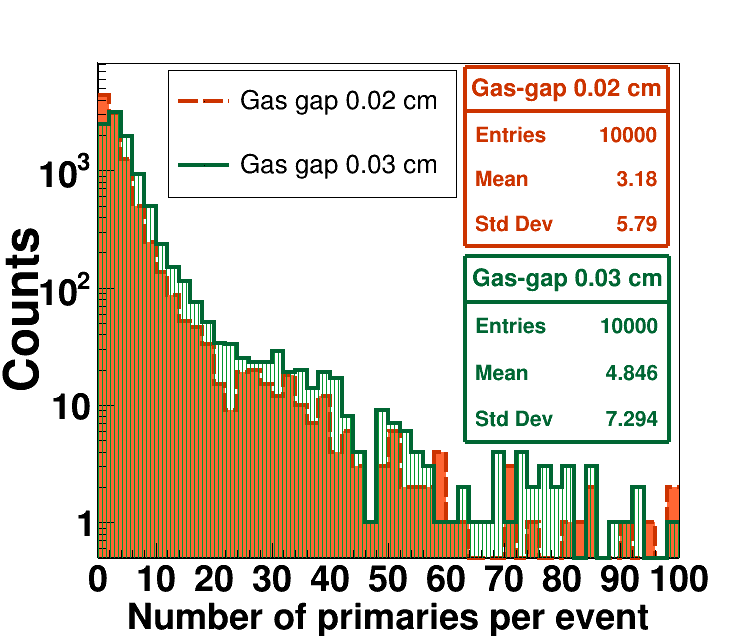}
			\caption{Disribution of number of primary electrons \label{fig:dist_prim}}
	\end{figure}

 Muon tracks of energy 2 GeV have been generated using HEED, built-in Garfield++. The direction of the tracks is chosen perpendicular to the surface of the RPC or along the positive z-direction. The distribution of primary electrons generated from a set of 10$^4$ muon tracks inside the gas-gap of three RPCs has been shown in Figure \ref{fig:dist_prim}, where the integral mean of the primary electron distribution for RPC1 and RPC2 is approximately 3, and 5 respectively. The prior figure shows that as the gap increased, the spectrum was also broadened, which is expected.
	
	\section{Induced charge distribution for different gas gap timing RPCs}\label{sec:indCh}
	A set of 10$^4$ avalanche is generated using the Montecarlo particle tracing model of Garfield++ inside the RPC1 and RPC2. The primary electrons are taken from the muon tracks mentioned in section \ref{sec:track}.
		  \begin{figure}[h]
		\center\includegraphics[scale=0.25]{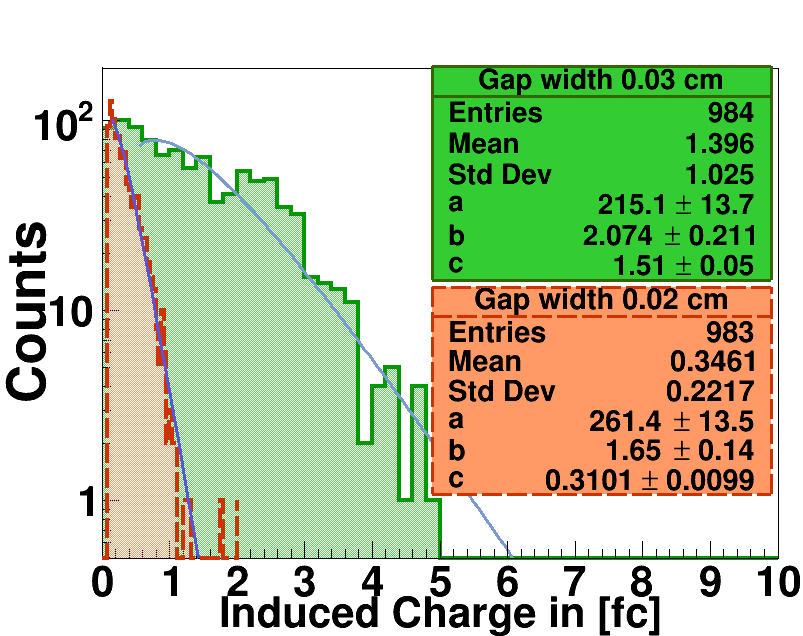}
		\caption{Disribution of number of primary electrons \label{fig:ind_chrg}}

	\end{figure}

   The induced charge $q^{ind}$ due to the movement of electrons and ions inside the RPC can be calculated by using Ramo's equation as follows \cite{ShockleyRamo}:
    \begin{equation}\label{eqn:induced_charge}
  q^{ind}=\int_{0}^{t} dt \sum_{n=0}^{N_{av}}\, \,q\,({W}^n({r_f}(t))-{W}^n({r_i}(t))),
  \end{equation}
  where ${W}^n({r_f}(t))$ and $\,{W}^n({r_i}(t)$ are the weighting potential at initial ($r_i$) and final position ($r_f$) of the step calculated by using neBEM. $N_{av}$ is the number of q point charges (electrons and ions) present in a step of time of the simulation.

 \par The induced charge distribution for RPC1 and RPC2 have been shown in Figure \ref{fig:ind_chrg}. In the prior figure, events are selected from the muon tracks, which contain a maximum of eight primaries, and a threshold of 0.1 fC on induced charge has been chosen. The charge distributions  of Figure \ref{fig:ind_chrg} have been fitted with a Polya function given as follows  \cite{polya_charge,KOBAYASHI2006136} :
   \begin{equation}
   f(q^{ind})=a(\frac{q^{ind}\,b}{c})^{b-1}\,e^{-\frac{b}{c}q^{ind}},
   \end{equation}
   where parameter ``a" is the scaling factor, ``b" is a free parameter that determines the shape of the distribution, and ``c" is the mean charge. From the values of fit
parameters of Figure \ref{fig:ind_chrg}, it can be said that the value of the parameter ``c” or mean charge increases with the increase in the gas gap, which is expected. For RPC1 and RPC2, the mean charge is 0.3 fC and 1.5 fC. The shape of the distribution is broadened with the increment of the gas gap, which reflects on the value of the parameter b as shown in Figure \ref{fig:ind_chrg}.

\section{Calculation of signal rise time} \label{sec:risetime}
   \begin{figure}[h]
  	\center\includegraphics[scale=0.25]{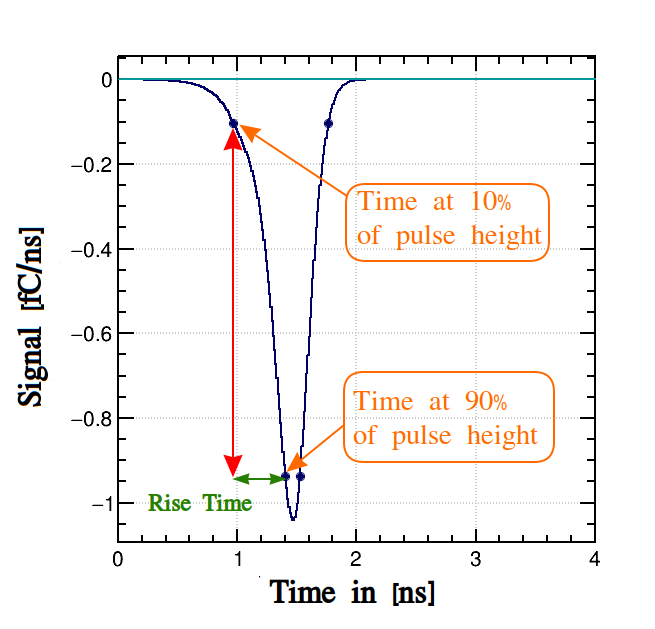}
  	\caption{Calculation of rise time of a signal. \label{fig:rise_time}}
  	
  \end{figure}
\begin{figure}[h]
	\center\includegraphics[scale=0.25]{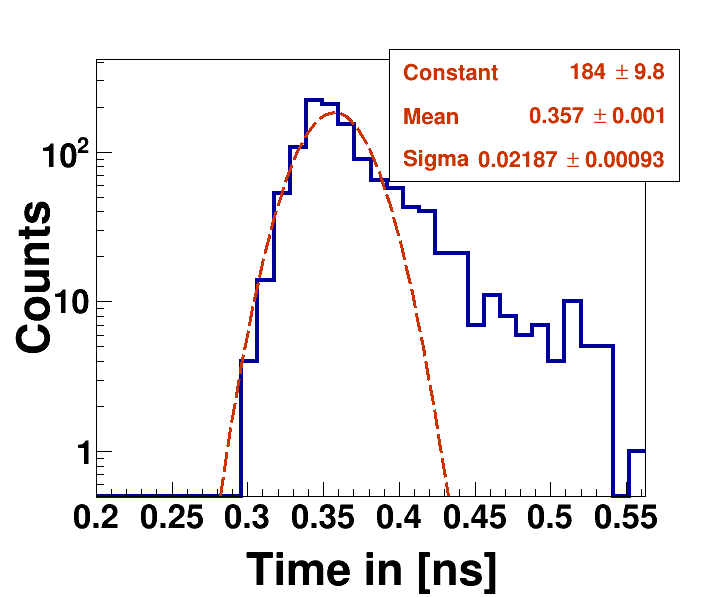}
	\caption{Rise time distribution of RPC1 \label{fig:time_reso_1}}
	
\end{figure}
\begin{figure}[h]
	\center\includegraphics[scale=0.25]{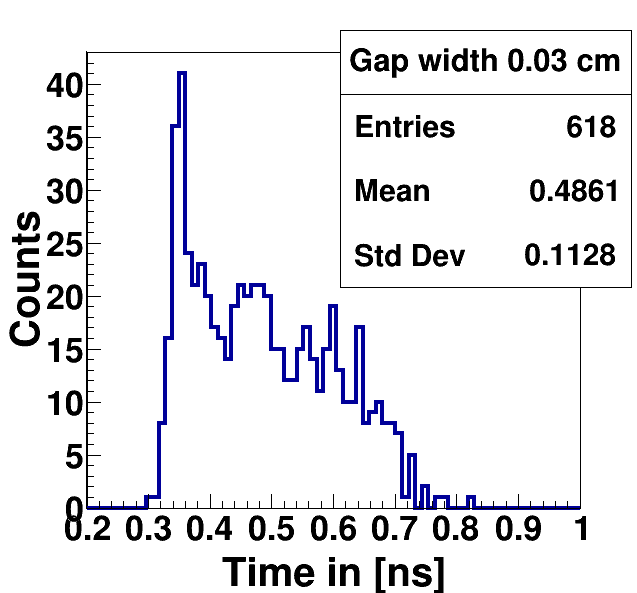}
	\caption{Rise time distribution of RPC2 \label{fig:time_reso_2}}
	
\end{figure}
\begin{figure}[h]
	\center\includegraphics[scale=0.25]{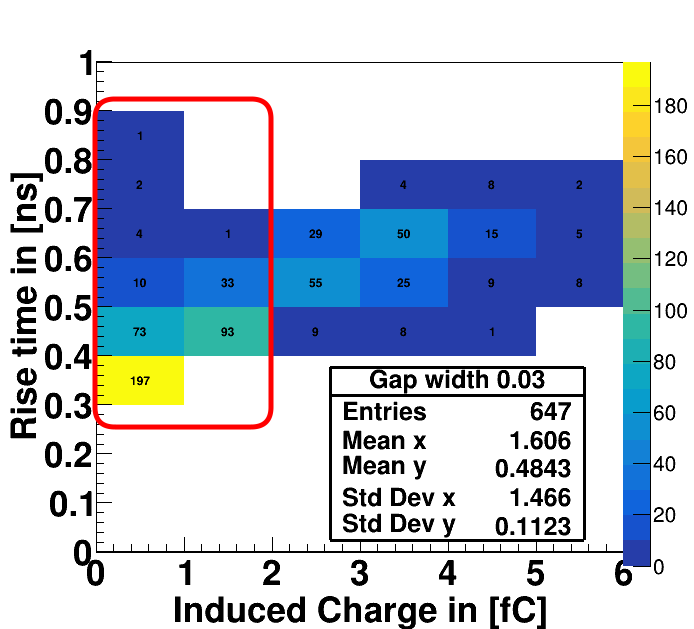}
	\caption{Correlation between Rise time and Induced charge of RPC2  \label{fig:time_reso_3}}
	
\end{figure}
\begin{figure}[h]
	\center\includegraphics[scale=0.25]{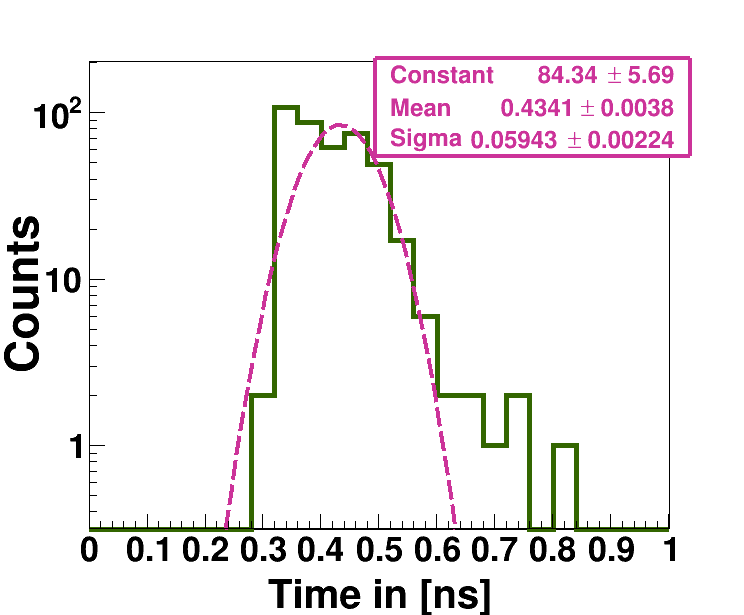}
	\caption{Rise time distribution of RPC2 for selected (< 2 fC) range of induced charge. \label{fig:time_reso_4}}
	
\end{figure}

The signal rise time can be calculated by taking the difference between the time at 90\% and 10\% of pulse height as shown in Figure \ref{fig:rise_time}. 
In Figure \ref{fig:time_reso_1} and \ref{fig:time_reso_2} the distribution of rise time for RPC1 and RPC2 has been shown. The rise time distribution of RPC1 (see Figure \ref{fig:time_reso_1}) is fitted with a gaussian function, and as a result, the time resolution or the sigma of the fit is approximately 21.8 ps.
 However, due to having multiple peaks in the rise time distribution of RPC2 (see Figure \ref{fig:time_reso_2}), the calculation of time resolution is not straightforward. Although, by looking at the correlation plot  (see Figure \ref{fig:time_reso_3}) of rising time vs induced charge and together with the mean induced charge value (1.5 fC) of RPC2, it can be said that below the 2 fC the probability of getting signal is more than other higher charge regions. Therefore, considering only those signals, the rise time distribution has been shown in Figure \ref{fig:time_reso_4}. The prior distribution has been fitted with a gaussian function. As a result of the sigma of the fit, it can be said that the time resolution of RPC2 is approximately 59.4 ps which is well matched with the experimental result shown in \cite{BLANCO200370}.  
 
 \section{Conclusion}
 In the present study, we have simulated the induced charge distribution and signal rise time distribution with the space charge effect of two different timing RPCs of gas gap 0.02 cm (RPC1) and 0.03 cm (RPC2).
 
 \par The induced charge distribution of RPC1 and RPC2 is fitted with a Polya function and found that the mean induced charge for RPC1 and RPC2 is 1.5 fC and 0.3 fC. Also found that the outline of the shape of the distribution is nonlinear due to the effect of the space charge, and the area of the induced charge region is more extensive in case of a larger gas gap.
 \par It is observed that as the gas gap increases, the timing performance of the RPCs becomes down due to the increment of time resolution and also, the mean rise time shifted towards the higher region. For the case of multiple peaks in the rise time distribution of RPC2, a cut on induced charge is introduced to select events and calculate time resolution. The simulation results of time resolution for 0.03 cm single gap RPC are verified with the experimental results.
 \par In the future, along with the RPCs, we will try to simulate avalanches in the other gaseous detectors with space charge effect and more detailed physics. A noise will be added to the signal to generate a more real signal.  
	\label{sec:8_summary}
\section{Acknowledgements}
All authors thank their respective universities and INO collaboration for their help and financial support.
	\bibliography{Image_issues_inRPC.bib}

\end{document}